\documentclass[notoc]{JHEP3}

\usepackage{graphicx}
\usepackage{dcolumn}
\usepackage{bm}
\usepackage{ulem}
\usepackage{amsmath}
\usepackage{psfrag}
\usepackage{epsfig}
\usepackage{wrapfig}

\newcommand*{\no}{\noindent}
\newcommand*{\bea}{\begin{eqnarray}}
\newcommand*{\eea}{\end{eqnarray}}
\newcommand*{\be}{\begin{equation}}
\newcommand*{\ee}{\end{equation}}

\newcommand*{\pref}[1]{(\ref{#1})}

\newcommand*{\mn}{{\mu\nu}}

\newcommand*{\nn}{\nonumber}

\newcommand*{\tr}{\mathrm{tr}}
\newcommand*{\diag}{\mathrm{diag}}

\renewcommand{\Re}{\mathrm{Re}\;}

\title{A first look at Landau-gauge propagators in G$_2$ Yang--Mills theory}

\author{Axel Maas\\ E-mail: \email{axelmaas@web.de}}
\author{\v{S}tefan Olejn{\'\i}k\\ E-mail: \email{stefan.olejnik@savba.sk}\\\\Department of Complex Physical Systems, Institute of Physics, Slovak Academy of Sciences, D\'{u}bravsk\'{a} cesta 9, SK-845 11 Bratislava, Slovakia}

\abstract{G$_2$ Yang--Mills theory is an interesting laboratory to investigate non-perturbative effects. On one hand, no conventional quark confinement via a linearly rising potential is present. On the other hand, its thermodynamic properties are similar to ordinary SU($N$) Yang--Mills theory. Finally, it has been conjectured that gluons are removed from the physical spectrum in the same way as in SU($N$) Yang--Mills theory. The last claim will be explored by determining the Landau-gauge ghost and gluon propagators, as well as the Faddeev--Popov operator eigenspectrum, in G$_2$ lattice gauge theory in two and three dimensions. The results are found to agree qualitatively with the SU(2) and SU(3) case. Therefore, the conjecture that Yang--Mills theories with different gauge groups are qualitatively similar on the level of their Landau gauge Green's functions is supported.}


\preprint{}

\begin{document}


\section{Introduction}

The experimental fact that neither quarks nor gluons, the elementary degrees of freedom of QCD, have been detected so far represents a theoretical challenge. In particular, experimental upper limits for the existence of free quarks are, due to their unique electromagnetic signature, impressively good \cite{Yao:2006px}, belonging to the class of the most precise of measurements to date.

The quest for a complete theoretical description of this phenomenon within the framework of QCD has not yet been entirely successful. However, there are two currently discussed mechanisms sufficient to make particles undetectable as individual entities in experiments. One of them is perfect screening, while the other one is that the quantum field does not at all represent a particle which belongs to the physical spectrum as an individual state.

The mechanism of perfect screening works as follows: Assume the potential between two infinitely heavy (i.e.\ static) test particles is confining, e.g.\ linearly rising like in the case of fundamental-representation quarks in pure SU($N$) Yang--Mills theory. As soon as the mass of test particles is finite, at some separation the rise of the potential produces enough energy for popping up a particle-antiparticle pair which then screens the test particles. In this way, a dynamical fundamental quark-antiquark pair cannot be separated arbitrarily far without generating new quarks which neutralize their color. Therefore, color is perfectly screened.  Such a potential, and also pair creation at sufficient separation, has been observed in lattice QCD \cite{Bali:2005fu}. Hence, quarks \textit{are} perfectly screened.

Note that an infinitely rising potential in the limit of infinitely massive particles is necessary for a perfect screening. Otherwise a sufficient amount of energy would permit to separate the constituents of a bound state to arbitrary distance, just as in the case of the ordinary electromagnetic Coulomb potential. Furthermore, it would be possible to thin out arbitrarily an ensemble of colored particles, like it is possible in case of an electromagnetic plasma, thereby isolating them on a macroscopic distance and make individual detection possible. Still, global color neutrality would hold, but not locally as in case of a bound state. This furthermore would entail questions on the definition of an almost localized non-Abelian charge, which is likely impossible \cite{Haag:1992hx}.

Perfectly screened particles can still be physical particles, although they do not have single particle asymptotic states\footnote{By this, the necessity to construct almost localized color charges \cite{Haag:1992hx} can be circumvented.}. Given a detector with sufficient spatial resolution, the particles could still be located inside bound states, e.g.\ in the case of quarks by their unique electromagnetic signature (fractional electric charge).

In the case of gluons it does not make much sense to speak about the potential between static infinitely heavy gluons. Nevertheless, gluons can still screen each other.

An alternative possibility to prevent particles from being detected individually is to remove them completely from the physical spectrum. This is, e.g., possible, if they do not have a K\"allen--Lehmann representation, which in turn is implied if the particles do not possess a positive definite spectral function. This has been observed explicitly for gluons (in Landau gauge) in lattice gauge theory \cite{Cucchieri:2004mf,Bowman:2007du} and in functional continuum calculations \cite{Alkofer:2003jj,Maas:2004se}. Therefore, gluons are not part of the physical spectrum, and are thus confined. Such particles could not be detected, not even in principle and with arbitrary spatial resolution, inside a bound state. In fact, the absence of colored particles from the spectrum implies a locally vanishing color density as an expectation value between physical states. Nonetheless, a hard gluon, dressed by soft gluons to make it color neutral, can still act as a current for spin, energy, and momentum, thereby manifest e.g.\ in jets.

Note that perfect screening and removal from the physical spectrum are not mutually exclusive. Both can apply to a particle at the same time.

Hence it seems to be clear -- up to the usual precautions on how these results have been obtained (see e.g.\ \cite{Aubin:2003ih,Aubin:2004av}) -- that both quarks and gluons are removed from the physical spectrum, explaining the experimental observation.

Still, this is only an empirical fact, and not yet an explanation of its dynamical origin, which is currently much less clear. Various topological degrees of freedom are suspected to be responsible for the generation of the confining potential, in particular vortices and monopoles are well-known alternatives \cite{Greensite:2003bk}. However, there are strong correlations between the various objects \cite{Greensite:2003bk,Ambjorn:1999ym,Boyko:2006ic,deForcrand:2000pg}, and a full and generally accepted understanding of the mechanism is still lacking, as is a complete understanding of its gauge dependence. For the removal of gluons from the physical spectrum the scenarios of Gribov and Zwanziger \cite{Gribov:1977wm,Zwanziger:1993dh,Zwanziger:2002ia,Zwanziger:2003cf} and of Kugo and Ojima \cite{Kugo:gm,Kugo:1995km} (generically denoted below as GZKO)  seem to be promising candidates. Their realization has been investigated in great detail for SU(2) and SU(3) Yang--Mills theory and QCD using functional methods  and in lattice gauge theory in various gauges, for overviews see \cite{Alkofer:2000wg,Fischer:rev}. However, there are still unresolved discrepancies between the results from various methods in particular in four dimensions \cite{Fischer:2007pf,Cucchieri:2007md,Sternbeck:2007ug,Bogolubsky:2007bw}, while in lower dimensional systems the agreement is qualitatively acceptable in three \cite{Maas:2004se,Cucchieri:2003di,Zwanziger:2001kw,Lerche:2002ep,Huber:2007kc} (however, see \cite{Cucchieri:2007md}) and even quantitatively excellent in two dimensions \cite{Maas:2004se,Zwanziger:2001kw,Lerche:2002ep,Huber:2007kc,Maas:2007uv}.

In addition, it is not yet known how the two aspects, confinement due to topological field configurations and the GZKO mechanism, fit together. Only recently first investigations have pointed to a deep relationship \cite{Greensite:2004bz,Langfeld:2002dd,Langfeld:2005kp,Maas:2005qt,Maas:2006ss}.

At this point Yang--Mills theory with gauge group G$_2$ becomes interesting. In contrast to ordinary SU($N$) Yang--Mills theory, a bunch of gluons can already screen a quark\footnote{It is always understood that a gluon is in the adjoint representation of the respective gauge group, while a quark is in the fundamental representation.} \cite{Holland:2003jy}. Hence perfect screening already occurs for infinitely heavy quarks, although the process of string-breaking implied by this has not yet been observed explicitly. Furthermore, due to the fact that the center of G$_2$ is trivial, the spectrum of topological configurations is qualitatively different from the one of SU($N$) Yang--Mills theory \cite{Greensite:2006sm}. However, other properties of SU($N$) Yang--Mills theory are retained, like a phase transition at finite temperature \cite{Greensite:2006sm,Pepe:2006er,Cossu:2007dk}. In particular, it has been conjectured, based on functional calculations, that gluons are removed from the physical spectrum by the GZKO mechanism, just as in SU($N$) Yang--Mills theory \cite{Maas:2005ym}. In this sense, G$_2$ Yang--Mills theory would be much more like SU($N$) QCD or SU($N$) Yang--Mills--Higgs theory rather than SU($N$) pure Yang--Mills theory. Therefore, G$_2$ Yang--Mills theory is an interesting laboratory to examine whether the ideas listed above on how quarks and gluons are prevented from being detected, are in fact correct.

The aim of the present pilot study is to subject the conjecture that gluons are removed from the physical spectrum and that this is operated by the GZKO mechanism also in G$_2$ Yang--Mills theory, to a first test. To this end, the Landau gauge propagators will be investigated. In particular, it will be checked whether the predictions of the GZKO mechanism for these propagators are correct. At this stage, this will be done in two and three dimensions for the sake of reducing computational requirements. The results are found to agree well with the ones in SU(2) and SU(3) Yang--Mills theory\footnote{We present thus the first results from lattice gauge theory on SU(3) Landau gauge propagators in two and three dimensions.}. Hence the conjecture of the GZKO mechanism to be at work in G$_2$ is supported. Furthermore, the removal of gluons from the physical spectrum is supported to the same extent as this is currently possible in SU(2) and SU(3) Yang--Mills theory using the information on the propagators alone for the given volumes. In addition, gauge fixing in G$_2$ Yang--Mills theory via stochastic overrelaxation is introduced in some detail.

The paper is organized as follows: In Section \ref{stech} technical aspects of the numerical simulations are described, in particular gauge-fixing in G$_2$ Yang--Mills theory. More in-depth technical details on the employed algorithms can be found in the appendix. Results on the gluon propagator will be presented in section \ref{sgluon}, while those on the ghost propagator in section \ref{sghost}. A final short summary will be given in section \ref{ssum}.

\section{Generation of gauge-fixed configurations}\label{stech}

\subsection{Configurations}

In the following, SU(2), SU(3), and G$_2$ Yang--Mills theory will be investigated in detail. If not specified otherwise, all statements apply equally well to Yang--Mills theory with all three gauge groups.

For the production of thermalized configurations, the standard Wilson action was employed, given by \cite{Montvay:1994cy}
\be
S=\beta\sum\left(1-\frac{1}{N_1}\Re\tr U_\mn\right)\nn,
\ee
\no where $N_1$ is the trace of the unit matrix in the fundamental representation, $\beta=2N_1/(g^2 a^{4-d})$, $d$ is the dimensionality, $a$ the lattice-spacing, $g$ the bare coupling, $U_\mn$ is the standard plaquette \cite{Montvay:1994cy}, and the sum is over all plaquettes. For SU(2), the Pauli representation for the links $U_\mu$ has been used, for SU(3) the Gell-Mann representation, and for G$_2$ the Macfarlane representation \cite{Macfarlane:2002hr} with explicit separation in the coset group G$_2$/SU(3) and the subgroup SU(3). $N_1$ takes then the values 2, 3 and 7, respectively.

\subsubsection{Setting the scale}

A significant problem in comparing the results for various gauge groups are the potentially different scales. This can be solved most easily by expressing all quantities by dimensionless ratios. In two and three dimensions this is particularly simple, as the gauge couplings have dimensions of energy (in two dimensions) and square root of energy (in three dimensions). Therefore by dividing all dimensionful quantities by appropriate powers of the gauge coupling, an explicit determination of the scale can be circumvented.

However, for lattice calculations an additional problem exists: An adequate determination of $\beta$ to obtain (approximately) the same value of the lattice spacing $a$ in physical units for all three gauge groups, and thus the possibility to do calculations in (approximately) the same physical volume. In two dimensions, where the string does not break in any Yang--Mills theory with arbitrary gauge group \cite{Dosch:1978jt}, the problem can be solved by direct determination of the fundamental asymptotic string tension\footnote{Alternatively, exact infinite-volume results, like for SU(2) gauge theory \cite{Dosch:1978jt}, could be used.}, which is a simple observable. In higher dimensions this is not possible in general: in G$_2$ the string breaks since fundamental G$_2$ charges can be screened by G$_2$ gluons. In three dimensions, instead, the string tension at intermediate distances can be used. The lattice volumes are anyhow not sufficiently large to detect the breaking of the string. 

\TABLE[h]{
\caption{\label{tconf}List of the configurations used. $d$ is the dimensionality and $N$ the size in lattice units. For the determination of $\beta$ and subsequently $a$, see text, and the dimensionful gauge coupling $g^2$, setting the natural scale. \textit{Sweeps} is the number of sweeps (each consisting of one heat-bath update and five overrelaxation updates) between two gauge-fixed measurements and \textit{cool} is the number of cooling sweeps from the (hot or cold) starting configuration until the first measurement. \textit{Plaquette} denotes the expectation value of the normalized plaquette. \textit{Threshold} is the maximum permitted value for the quantity $e_6$, defined in \pref{e6}, after gauge fixing. $p$ is the threshold value for the stochastic overrelaxation \cite{Cucchieri:1995pn}, which has been determined using an adaptive algorithm \cite{Cucchieri:2006tf}.}
\begin{tabular}{|c|c|c|c|c|c|c|}
\hline
$d$/Group & 2d/SU(2) & 2d/SU(3) & 2d/G$_2$ & 3d/SU(2) & 3d/SU(3) & 3d/G$_2$ \\
\hline
$N$ & 34 & 34 & 34 & 16 & 16 & 16 \\
\hline
$\beta$ & 10 & 28 & 50 & 4.24 & 10.7 & 18.8 \\
\hline
$1/a$ [GeV] & 1.10 & 1.10 & 1.10 & 1.14 & 1.14 & 1.14 \\
\hline
$V^{1/d}$ [fm] & 6.09 & 6.09 & 6.09 & 2.76 & 2.76 & 2.76 \\
\hline
$ag^{d-1}$ & 0.632 & 0.463 & 0.529 & 0.943 & 0.561 & 0.745 \\
\hline
Cool & 440 & 440 & 440 & 360 & 360 & 360 \\
\hline
Sweeps & 44 & 44 & 44 & 36 & 36 & 36 \\
\hline
Conf. & 1038 & 1756 & 166 & 1048 & 985 & 149 \\
\hline
Plaquette & 0.85412(2) & 0.85865(1) & 0.86010(2) & 0.74456(2) & 0.72483(1) & 0.72397(2) \\
\hline
Threshold & $10^{-12}$ & $10^{-11}$ & $10^{-11}$ & $10^{-12}$ & $10^{-11}$ & $10^{-11}$ \\
\hline
$p$ & 0.79 & 0.79 & 0.79 & 0.58 & 0.55 & 0.46 \\
\hline
\end{tabular}
}

\EPSFIGURE[h]{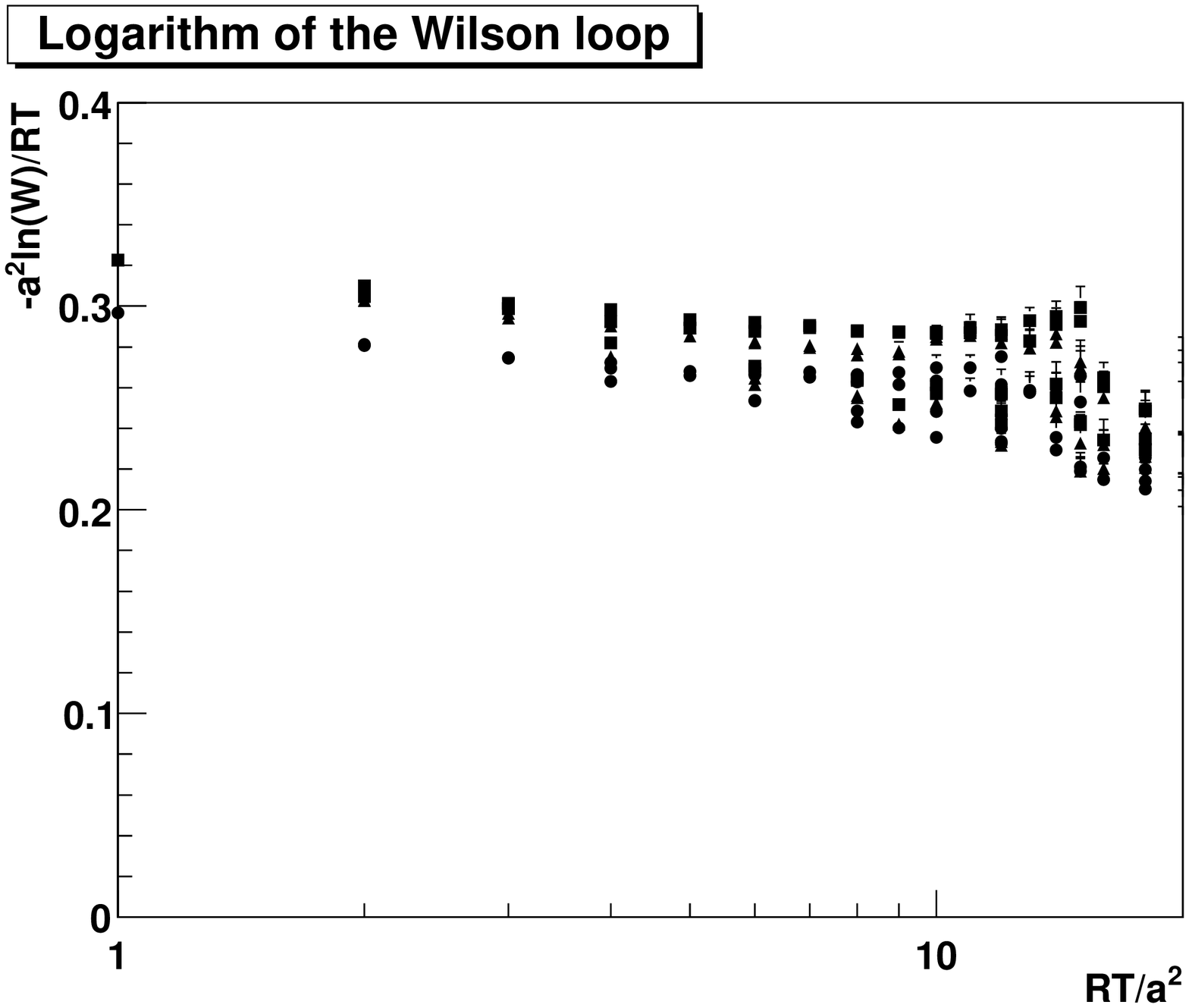,width=0.75\linewidth}{\label{fwilson}The logarithm of the Wilson loop W divided by the area RT of the loop, in three dimensions, obtained from the systems listed in table \ref{tconf}. Circles are from SU(2), squares are from SU(3), and triangles are from G$_2$ Yang-Mills theory. All possible values for the spatial extent R and the temporal extent T of the loop for a given area RT have been included.}

Unfortunately, even a direct measurement of the intermediate distance string tension is difficult and nontrivial in three dimensions, because various corrections to the string tension are present. For SU(2) and SU(3), the values\footnote{Interpolated, where necessary.} from \cite{Teper:1998te} will be used. For G$_2$ similar results are not available. However, it turns out that the gluon and ghost propagator show scaling within statistical errors when the value of $a$ is determined according to the following procedure: First, the ratio $r^{d=2}$ of $\beta$ values is determined between SU(4) and G$_2$ giving the same string tension $\sigma$ in two dimensions,
\be
r^{d=2}=\frac{\beta_{\mathrm{SU(4)}}^{d=2}(\sigma)}{\beta_{\mathrm{G}_2}^{d=2}(\sigma)}\nn.
\ee 
\no Using the known $\beta$-string tension relation for SU(4) in three dimensions \cite{Teper:1998te}, the corresponding $\beta$-value for a desired string tension in three dimensions is obtained. The $\beta$-values for SU(4) in three dimensions are then scaled by the factor $r^{d=2}$ determined in two dimensions to obtain $\beta$-values for G$_2$,
\be
\beta_{\mathrm{G}_2}^{d=3}=r^{d=2}\;\beta_{\mathrm{SU(4)}}^{d=3}\nn.
\ee 
\no Then, $a$ for G$_2$ is set to the same value as it has for this string tension in SU(4). The resulting $\beta$-values are given in table~\ref{tconf}. It is quite interesting that this procedure seems to work reasonably well. It hints that in Yang--Mills theory the $\beta$-$a$-relation is essentially dominated by the number of generators, which are similar for G$_2$ (14) and SU(4) (15). Interestingly enough, a-posteriori this procedure is justified by looking at the behavior of the logarithm of the Wilson loop divided by the loop area, versus the area in physical units \cite{Montvay:1994cy}. The comparison for various gauge groups is shown in figure \ref{fwilson}. The resulting values (that approach the string tension for large enough loop areas) are for all gauge groups essentially indistinguishable, and, therefore, provided that the finite volume/spacing correction terms are not adjusting themselves such as to create accidentally equality, the intermediate string tensions in physical units are approximately the same. Hence this procedure really permits to select the $\beta$-values such that the physical volumes are approximately the same.

Therefore, it would then be possible to obtain the results in units of $a$ or in units of the string tension\footnote{This is equivalent to assign the string tension in all cases the same value in units of GeV, e.g.\ the quasi-standard value $\sqrt{\sigma}=$ 440~MeV.}. Comparing the various results once in units of the respective coupling constants and once in units of $a$, it is immediately visible that the quantitative changes are surprisingly strong. This is already the case between SU(2) and SU(3), where no complications in determining the string tension arise. For example, the gluon propagator is essentially identical when measured in units of the string tension, while it is almost a factor of two different in the infrared when measured in units of the coupling constants. The situation is reversed for the ghost propagator, it is more similar when given in units of the coupling constant. However, this property does not pertain to the spectrum of the Faddeev-Popov operator itself\footnote{The inversion step from the Faddeev-Popov operator to the ghost propagator mixes all scales.}. Therefore, a quantitative comparison of the results for the different gauge groups is not sensible, and only qualitative features will be discussed.

Furthermore, here the only matter of interest is, whether at all a similarity between the gluonic correlation functions between G$_2$ and SU($N$) gauge groups can be expected. A detailed determination of the volume dependence is thus left to the future. The results below will be given in units of the coupling constants.

\subsubsection{Generation of thermalized configurations}

Configurations are then obtained by a mix of heat-bath and overrelaxation sweeps. This has been done for SU(2) as described in \cite{Cucchieri:2006tf}. In SU(3), heat-bath updates have been performed using the Cabibbo-Marinari method \cite{Cabibbo:1982zn} with three SU(2) subgroups, and overrelaxation with five overrelaxation sweeps between two heat-bath updates. In the case of G$_2$, heat-bath updates have been performed according to the method presented in \cite{Pepe:2006er} and detailed in \cite{Greensite:2006sm}. For this purpose, after an ordinary heat-bath update of an SU(3) subgroup, a random gauge-transformation was performed to update another subgroup. This was repeated five times. In between two heat-bath sweeps, five over-relaxation sweeps were done, each again with five random gauge transformation mixes. An overrelaxation step was performed by overrelaxing an SU(2) subgroup, which was selected using a randomly generated similarity transformation from the coset group G$_2$/SU(3), see appendix \ref{aover} for details. This was repeated for three SU(2) subgroups. Including the random gauge transformation sweeps therefore 15 SU(2) subgroups are subjected to each overrelaxation update. However, these numbers could of course be optimized which was not attempted in this pilot study.

Note that here random G$_2$ matrices, either in the gauge transformations or elsewhere, are always generated anew instead from a lookup table \cite{Pepe:2006er,Greensite:2006sm}. For this purpose, elements from the SU(3) subgroup have been generated randomly by generating three elements of the three SU(2) subgroups randomly to construct one SU(3) element, while elements from the coset group G$_2$/SU(3) are generated from a complex three dimensional vector with elements drawn from a normal distribution with unit width. This distribution was used instead of a uniform distribution \cite{Greensite:2006sm} because coset elements from vectors with large elements improve mixing, but too large elements slow down equilibration.

For the present purpose of a first look at qualitative features, only one single lattice volume in two and three dimensions will be considered. The details of the configurations used are given in table \ref{tconf}. Note that due to the increase in the number of generators fewer configurations are necessary for SU(3) than for SU(2), and for G$_2$ than for SU(3) to reach the same level of statistical accuracy for color-averaged gluonic observables in general. However, single exceptional configurations can alter these requirements for a finite amount of statistics.

\subsection{Gauge-fixing}

Once an equilibrated configuration is obtained, it is still necessary to fix it to the Landau gauge. This is done using a stochastic overrelaxation algorithm by minimizing the functional (see, e.g., \cite{Cucchieri:1995pn})
\be
{\cal E}=-\sum\Re\tr U_\mu,\nn
\ee
\no where $\{ U_{\mu}(x) \}$ is a thermalized lattice configuration, and the sum is over all links. This leads to the so-called minimal Landau gauge. For SU(2), this is done using stochastic overrelaxation with adaptive parameter adjustment \cite{Cucchieri:2006tf}. For the other gauge groups, the standard overrelaxation step has to be modified compared to the one in SU(2). This can be done by over-relaxing all SU(2) subgroups for both, SU(3) and G$_2$, in the same manner as for overrelaxation sweeps during the generation of configurations, in particular the same number of subgroups. See again appendix \ref{aover} for the case of G$_2$. However, in the case of G$_2$, of course, the random gauge transformations are replaced by random similarity transformations of the gauge transformation to select randomly an SU(2) subgroup.

Still, it is necessary to project the resulting matrices back into the group. This is trivial for SU(2), and done according to \cite{Suman:1993mg} for SU(3). For G$_2$, once more, a similar strategy as for SU(3) can be applied. However, the projection back into G$_2$ is more complicated, than in the case of SU(3). It is performed by constructing the projected matrix as a product of projected SU(3) subgroups of G$_2$. This algorithm is detailed in appendix \ref{aproj}. 

This procedure is sufficient for gauge-fixing. Its quality is monitored using the quantity $e_6$, defined as \cite{Cucchieri:1995pn}
\bea
e_6&=&\frac{1}{d}\sum_\mu\frac{1}{N_\mu}\sum_c\frac{1}{[\tr(Q_\mu t_c)]^2}\label{e6}\\
&&\quad\qquad\times\sum_{x_\mu}(\tr\{[q_\mu(x_\mu)-Q_\mu]t_c\})^2\nn\\
q_\mu(x_\mu)&=&\frac{1}{2i}\sum_{x_\nu,\nu\neq\mu}\big[g(x)U_\mu(x)g(x+e_\mu)^\dagger\nn\\
&&\qquad\qquad -g(x+e_\mu)U_\mu(x)^\dagger g(x)^\dagger \big] \nn \\
Q_\mu&=&\frac{1}{N_\mu}\sum_{x_\mu}q_\mu(x_\mu)\nn,
\eea
\no where  $\{ g (x) \}$ represents the gauge transformation applied on the link variables $ U_{\mu}(x) $, the symbol $^\dagger$ indicates Hermitian conjugation, $ N_{\mu} $ is the lattice side in the $\mu$ direction, $d$ is the space-time dimensionality, $e_{\mu}$ is a positive unit vector in the $\mu$ direction and $t_{c}$ are the generators of the algebra. This quantity is a more reliable measure of the gauge fixing quality than just the transversality itself \cite{Cucchieri:1995pn}. Furthermore, it is found that the same limit of $e_6$ corresponds to a much better fulfillment of the transversality condition with increasing number of generators. Therefore, the restriction on $e_6$ for achieving the gauge-fixing is and can be taken somewhat lower for G$_2$ and SU(3) than for SU(2), corresponding still to a better level of transversality on the average. The actual values are given in table \ref{tconf}. Note that in general a level of 10$^{-9}$ of $e_6$ is already sufficient for the propagators not to change anymore within statistical errors for the number of configurations used here. In the case of a gauge-fixing worse than that level the effects are quickly visible and most pronounced at larger momenta in the gluon propagator, and show up very strongly in the ghost propagator at all momenta, due to its character as the expectation value of an inverted operator.

With these methods, the gauge-fixed configurations have been obtained. No attempt is made in this first investigation to reduce or estimate the effects of Gribov copies, although, at least at the rather small volumes used here, some effects are to be expected, as is known from investigations in four dimensions\footnote{Note that the center-flips advocated in \cite{Bogolubsky:2007bw} to improve the gauge-fixing of course cannot be implemented in the case of G$_2$, as the center is trivial.} \cite{Cucchieri:1997dx,Bogolubsky:2007bw}. This has to be investigated in detail in the future. Finally, it should be remarked that this gauge-fixing procedure for G$_2$ can be trivially adapted to also fix the minimal Coulomb gauge or gauges interpolating between the Landau and Coulomb gauge.

\section{The gluon propagator}\label{sgluon}

The color-averaged, scalar gluon propagator $D(p)$ is defined in the usual way as
\bea
D(p)&=&\delta_{ab}\left(\delta_\mn-\frac{p_\mu p_\nu}{p^2}\right)D_\mn^{ab}(p)\nn\\
D_\mn^{ab}(p)&=&\frac{1}{V}\langle A_\mu^a(p)A_\nu^b(-p)\rangle,\nn
\eea
\no where momenta $p$ are the standard lattice momenta. The expectation from functional calculations \cite{Alkofer:2000wg,Fischer:rev,Lerche:2002ep,Huber:2007kc,Zwanziger:2001kw,Alkofer:2004it,Fischer:2006vf,vonSmekal:1997is,Pawlowski:2003hq} is that the scalar gluon propagator $D(p)$ follows a power law in the far infrared, with a characteristic exponent $t$, $D(p)\sim p^{-2-2t}$. The value for this exponent is not uniquely determined in functional calculations due to uncertainties in the required truncations. However, with assumptions on the vertices compatible with lattice results \cite{Cucchieri:2006tf,Maas:2007uv,Cucchieri:2004sq,Ilgenfritz:2006he} and self-consistency tests in functional calculations \cite{Schleifenbaum:2004id}, $t$ should be smaller than $-1$ for all dimensions, leading to an infrared-vanishing gluon propagator.

In particular, a gluon propagator which vanishes at zero momentum implies necessarily positivity violation and the gluon would be removed from the physical spectrum. However, this can only occur in an infinite volume, and corresponding extrapolations are necessary. Here, just a check on a finite volume is performed. Only by assuming that a similar propagator for all groups implies a similar behavior in an infinite volume, a conjecture at this point can be made on the removal of gluons from the spectrum. Further studies on the evolution with volumes are therefore required for a substantial statement. However, the aim here is only to test whether it is reasonable at all to expect a similar evolution for all the different gauge groups.

\TABLE[h]{
\caption{\label{tpert}Leading order perturbative contributions in three dimensions \cite{Maas:2004se}. Note that no resummation at this order takes place in three dimensions. $C_A$ is the adjoint Casimir, and normalized according to the conventions of \cite{Alkofer:2000wg}. The value for the $g^2$ are taken from the $\beta$-values of the simulation as $g^2=2N_1/(\beta a^{4-d})$.}
\vspace{1mm}
\begin{tabular}{|c|c|c|c|c|}
\hline
Group & $ag^2$ & $C_A$ & Gluon propagator & Ghost propagator \\
\hline
SU(2) & 0.943 & 2 & $\frac{11g^2 C_A}{64p}=\frac{0.324}{ap}$ & $\frac{g^2 C_A}{16p}=\frac{0.118}{ap}$ \\
\hline
SU(3) & 0.564 & 3 & $\frac{0.291}{ap}$ & $\frac{0.108}{ap}$ \\
\hline
G$_2$ & 0.745 & 2 & $\frac{0.256}{ap}$ & $\frac{0.0931}{ap}$ \\
\hline
\end{tabular}}

Note that neither in two, nor in three dimensions any physical renormalization occurs, although lattice artifacts may produce additional effects \cite{Lepage:1992xa}. Already in perturbation theory the gluon (and ghost) propagator differs for the different groups. E.g., the leading order perturbative coefficient changes by the ratio of the adjoint Casimir operator from gauge group to gauge group. Nonetheless, this effect is rather small, as at momenta where perturbation theory is applicable, the effects are anyway almost negligible compared to the dominating tree-level values. In particular, the dimensionful coupling, which changes for each gauge group with the $\beta$-values employed here, and which is not renormalized, enters as well, compensating the increase with the Casimir at least partly. Therefore, these effects on the perturbative tail are in fact not directly visible in the result from lattice calculations presented here. For three dimensions, the explicit values are given in table \ref{tpert}. At the present values of $g$, the corrections are already essentially negligible, when perturbation theory can be applied.

To investigate only the color-diagonal elements is justified, as the off-diagonal elements vanish for sufficiently large volumes and statistics. This has been checked explicitly here. In case of SU(2) this may not yet be surprising, as the color-symmetric structure constants vanish. However, it seems so far also to apply to SU(3) and G$_2$ as a gauge group, although their symmetric structure constants are no longer zero. This is in agreement with previous studies of these off-diagonal components in two and three dimensions for SU(2) \cite{Cucchieri:2006tf,Maas:2007uv}.

\FIGURE[h]{
\includegraphics[width=0.5\linewidth]{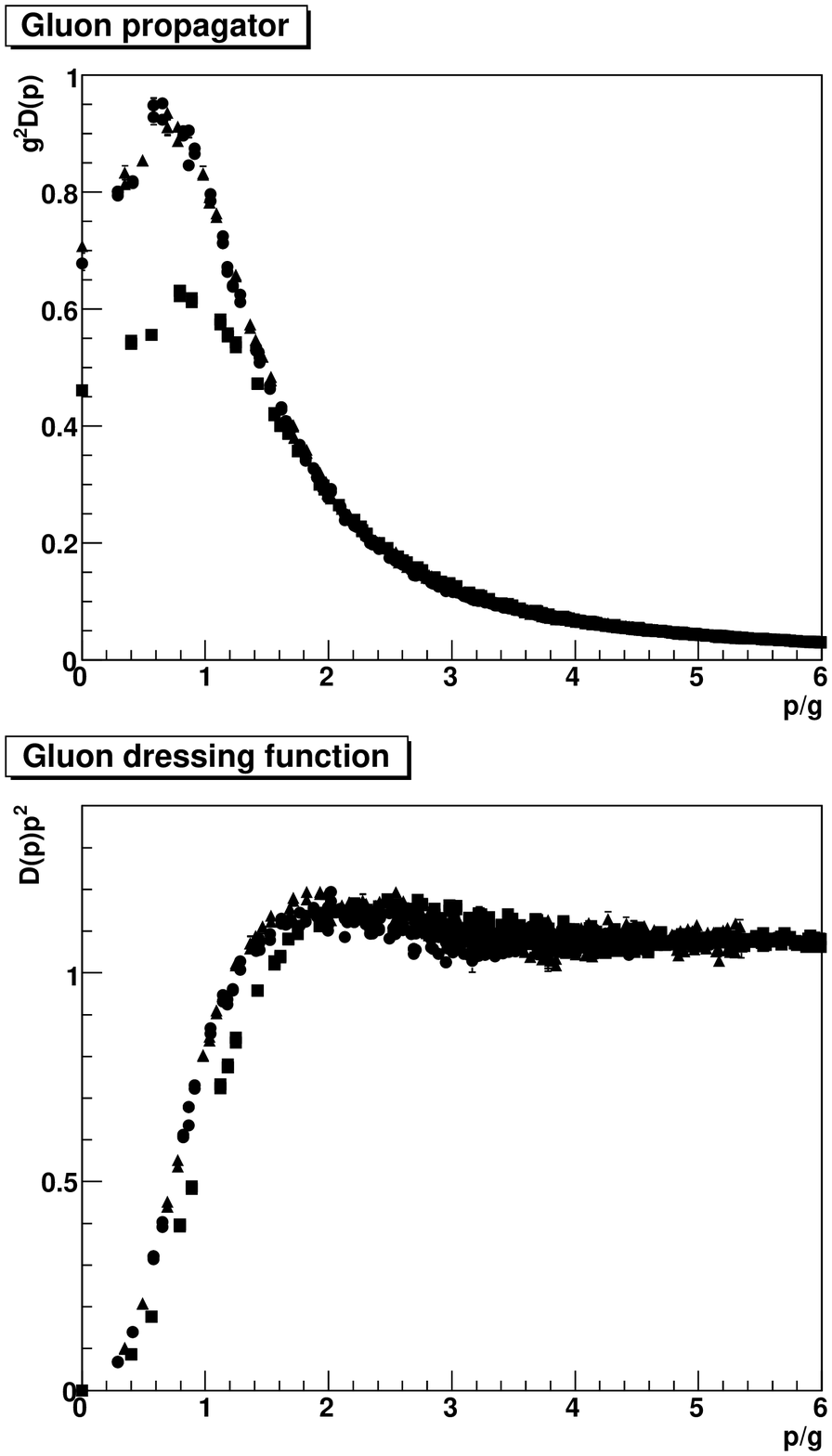}\includegraphics[width=0.5\linewidth]{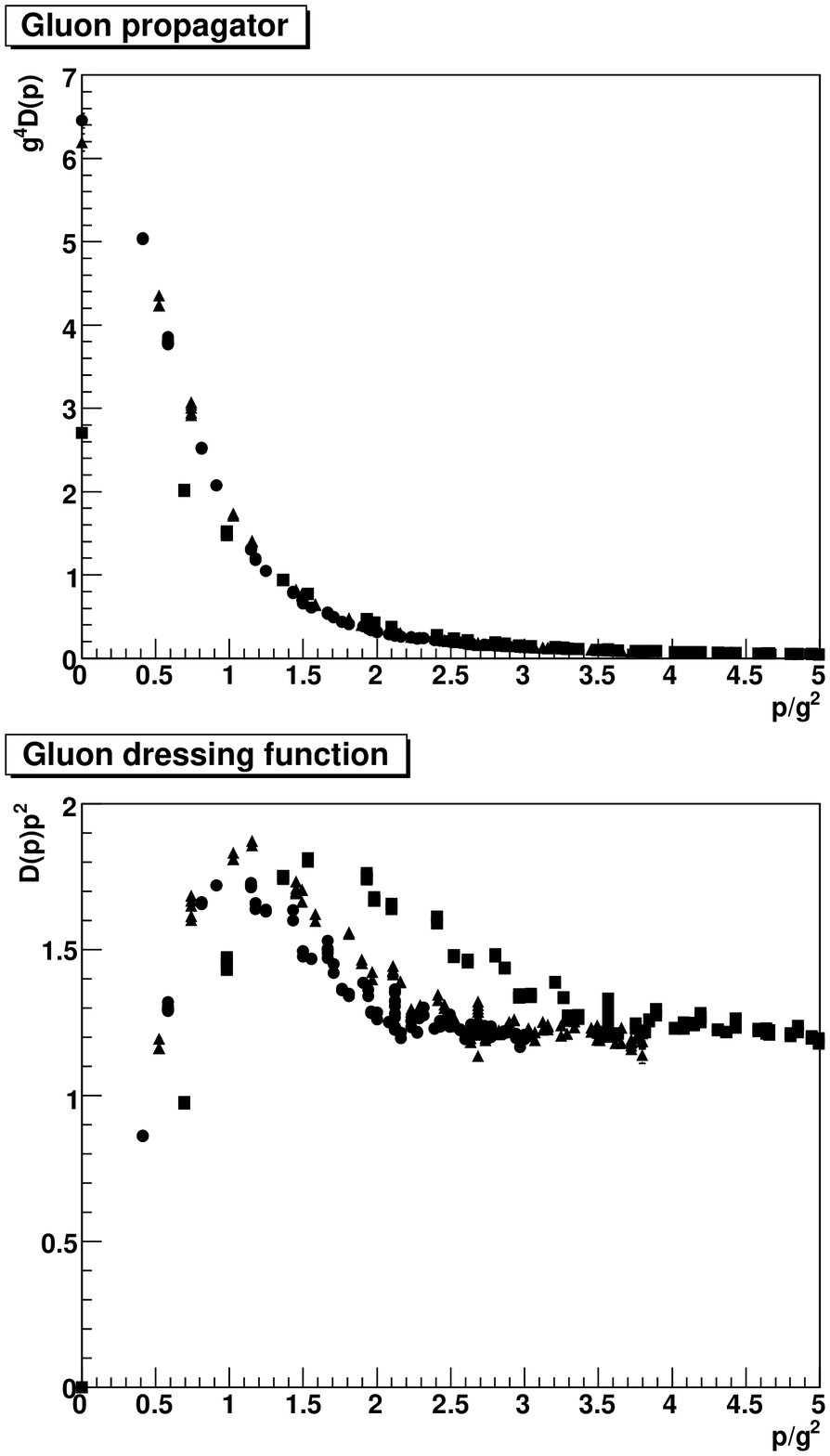}
\caption{\label{fgp}The gluon propagator (top panels) and the gluon dressing function (lower panels) in two (left panels) and three (right panels) dimensions. Dimensionful quantities have been normalized by the gauge coupling to obtain dimensionless quantities. Circles are from SU(2) Yang--Mills theory, squares are from SU(3) Yang--Mills theory, and triangles are from G$_2$ Yang--Mills theory. Values for momenta from edge, arbitrary plane and volume-diagonal (in three dimensions) momentum configurations are plotted \cite{Cucchieri:2006tf}, but not marked differently. Violation of rotational symmetry is thus relatively small.}}

The results are shown for two and three dimensions in figure \ref{fgp}. In two dimensions, in all cases, the gluon propagator is exhibiting a clear maximum, and is infrared suppressed. Also at larger momenta there is no qualitative difference for the various gauge groups. Even the relative infrared suppression compared to the height of the respective maximum is quite similar at this similar volume.

In three dimensions, none of the propagators exhibits (yet) a maximum. In the case of SU(2) it is known that this will happen eventually for large enough volumes. However, the properties of the propagators at this volume are qualitatively the same for all gauge groups. Assuming that a similar evolution with volume exists in all cases, also the propagators for SU(3) and G$_2$ will turn over at sufficiently large volumes, just as in two dimensions.

The common property in all cases is that the propagator bends over in the far infrared in two dimensions. In three dimensions the volumes are not yet large enough to observe this. To which extent these are quantitatively similar in the infinite volume limit, is open. In particular, it could happen that the infrared exponent of the gluon propagator depends on the gauge group. In two dimensions, it is possible to extract it \cite{Maas:2007uv}, but larger volumes than the current ones will be necessary for this purpose. Such a dependence would be at variance with the expectation due to functional studies \cite{Schleifenbaum:2004id}. However, it would still be consistent, if the ghost-gluon vertex exhibited a different behavior for different gauge groups. At least for small volumes, this seems not to be the case in the comparison of SU(2) and SU(3) Yang--Mills theory in four dimensions \cite{Cucchieri:2004sq,Ilgenfritz:2006he}.

Nonetheless, in all three cases the same qualitative behavior is found. Hence, for all three gauge groups the gluons are removed likely in the same way from the physical spectrum. As stated above, this conjecture is based only on a similar behavior of all propagators at a fixed lattice system, and will require further investigations to substantiate.

It is worthwhile to add some remarks on the comparison in four dimensions \cite{Cucchieri:2007zm}. The difference at small momenta between SU(2) and SU(3) is more pronounced in two and three dimensions quantitatively. On the other hand, the difference at large momenta is less pronounced, since the ratio between both propagators goes to infinity in four dimensions and to one in three or two dimensions, due to renormalization effects. The relevant scale, which influences the infrared behavior, evolves in four dimensions by dimensional transmutation in the renormalization process. In three or two dimensions, this scale is set explicitly by the value of $g$. This influences of course the ratio of the propagators, as the respective values of $g$ are different for the different groups. This qualitative difference of how the scale is set in the different dimensions can cause the difference in the ratio of propagators.

\section{The ghost propagator}\label{sghost}

The ghost propagator is defined as the inverse of the Faddeev--Popov operator $M^{ab}$. $M^{ab}$ is defined in the same way for all gauge groups by its action on a function on the lattice as \cite{Zwanziger:1993dh}
\bea
M(y,x)^{ab}\omega_b(x)&=&c\left(\sum_x\big(G^{ab}(x)\omega_b(x)+\sum_\mu A_\mu^{ab}(x)\omega_b(x+e_\mu)+B_\mu^{ab}(x)\omega_b(x-e_\mu)\big)\right)\nn\\
G^{ab}(x)&=&\sum_\mu \tr(\{t^a,t^b\}(U_\mu(x)+U_\mu(x-e_\mu)))\nn\\
A_\mu^{ab}(x)&=&-2\tr(t^a t^bU_\mu(x))\nn\\
B_\mu^{ab}(x)&=&-2\tr(t^a t^bU_\mu^{+}(x-e_\mu))\nn,
\eea
\no where $c$ is a constant depending on the normalization of the generators $t^a$. Again, the expectation from functional calculations \cite{Alkofer:2000wg,Fischer:rev,Lerche:2002ep,Huber:2007kc,Zwanziger:2001kw,Alkofer:2004it,Fischer:2006vf,vonSmekal:1997is,Pawlowski:2003hq} is that the ghost propagator $D(p)$ follows a power law in the far infrared, with a characteristic exponent $\kappa$, $D_G(p)\sim p^{-2-2\kappa}$. The GZKO mechanism only implies that $\kappa$ should be larger than zero. Functional methods, under additional assumptions, lead to explicit values and relations to the exponent $t$ of the gluon propagator. However, here only the qualitative statement of the GZKO mechanism that $\kappa$ should be larger than zero will be tested.

The numerical inversion process, a conjugation gradient inversion \cite{Cucchieri:2006tf}, is the same for all gauge groups, and thus needs not to be changed. Again, only the color-averaged color-diagonal part will be investigated, which is the only one different from zero for sufficiently high statistics for SU(2) \cite{Maas:2007uv,Cucchieri:2006tf}. This does not change for the gauge group SU(3), for which this was known in four dimensions \cite{Boucaud:1998xi}, and it turns out that this also applies to G$_2$.

\FIGURE[h]{
\includegraphics[width=0.5\linewidth]{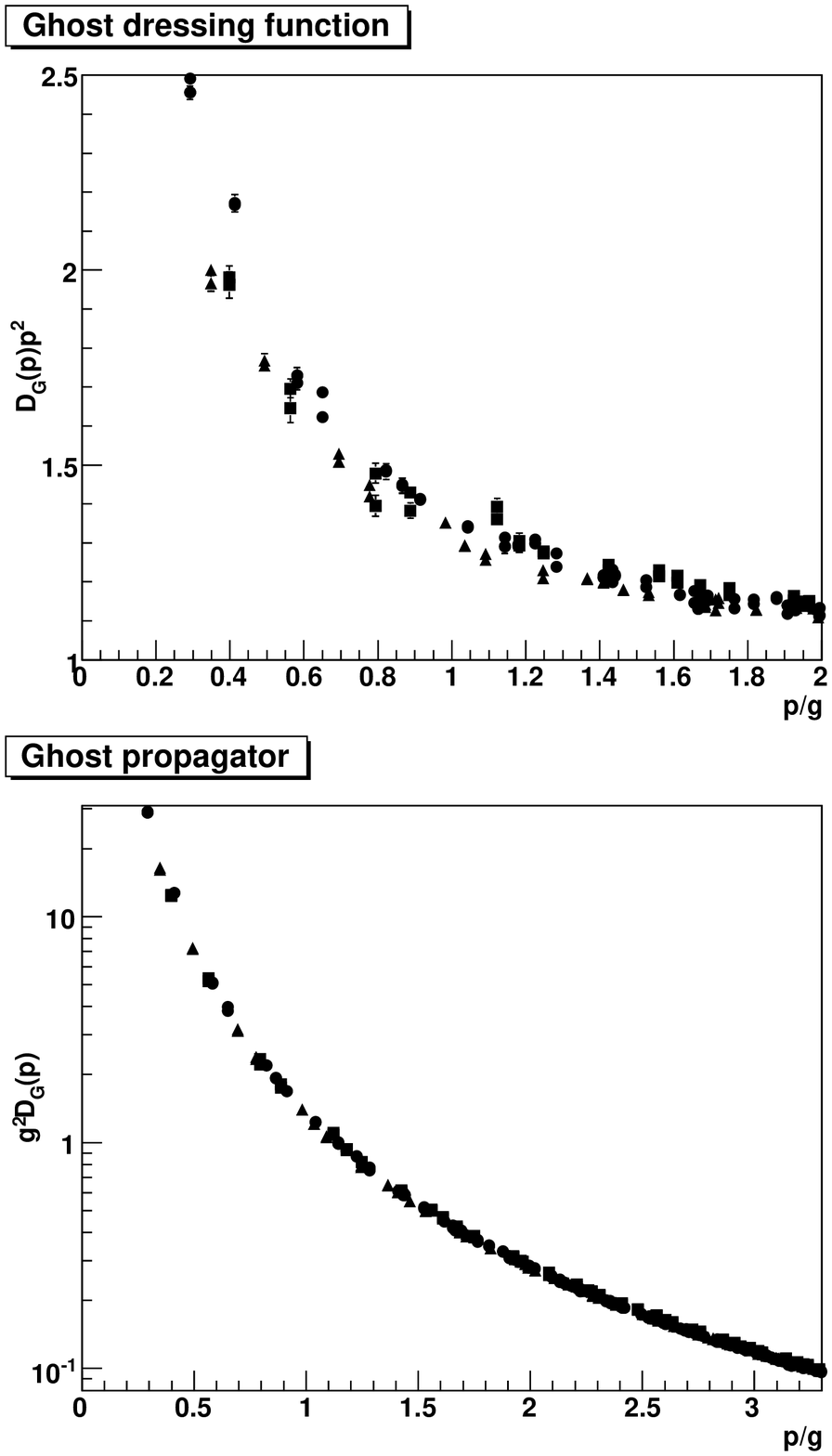}\includegraphics[width=0.5\linewidth]{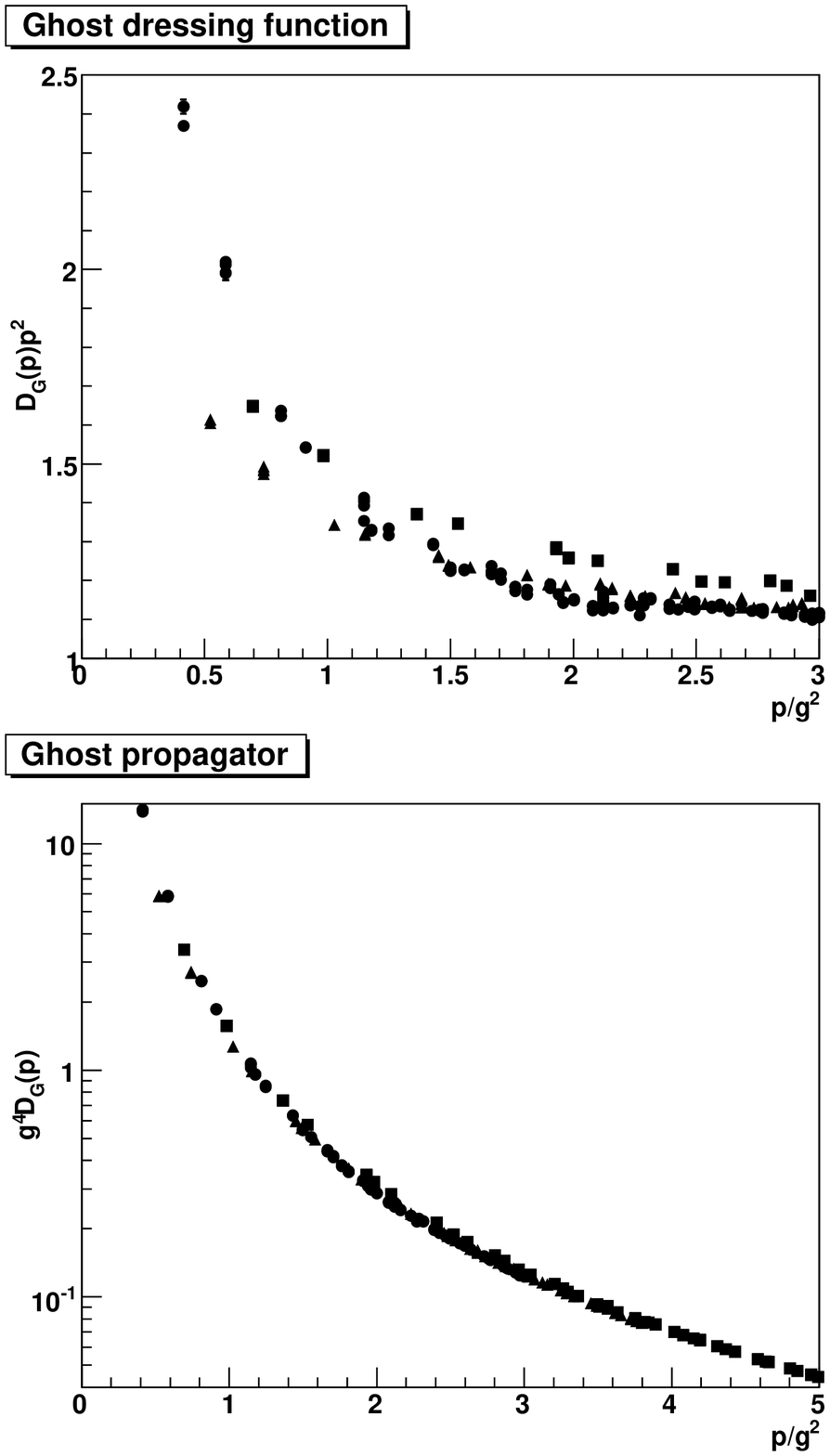}
\caption{\label{fghp}The ghost propagator (top panels) and the ghost dressing function (lower panels) in two (left panels) and three (right panels) dimensions. Circles are from SU(2) Yang--Mills theory, squares are from SU(3) Yang--Mills theory, and triangles are from G$_2$ Yang--Mills theory. Momenta from edge, arbitrary plane and volume-diagonal (in three dimensions) momentum configurations \cite{Cucchieri:2006tf} are plotted, but not marked differently. Violation of rotational symmetry is thus relatively small.}}

\FIGURE[h]{
\includegraphics[width=0.5\linewidth]{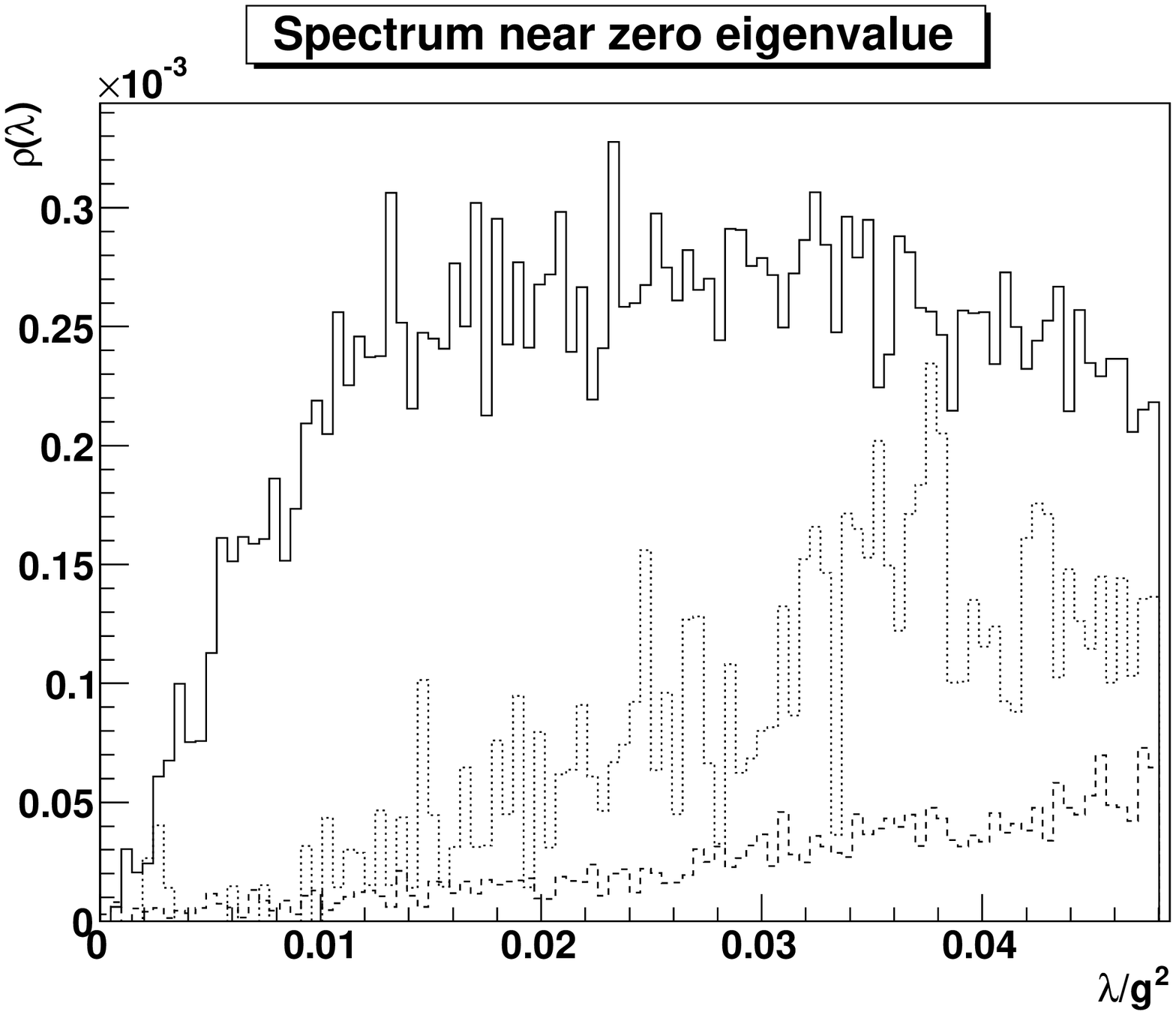}\includegraphics[width=0.5\linewidth]{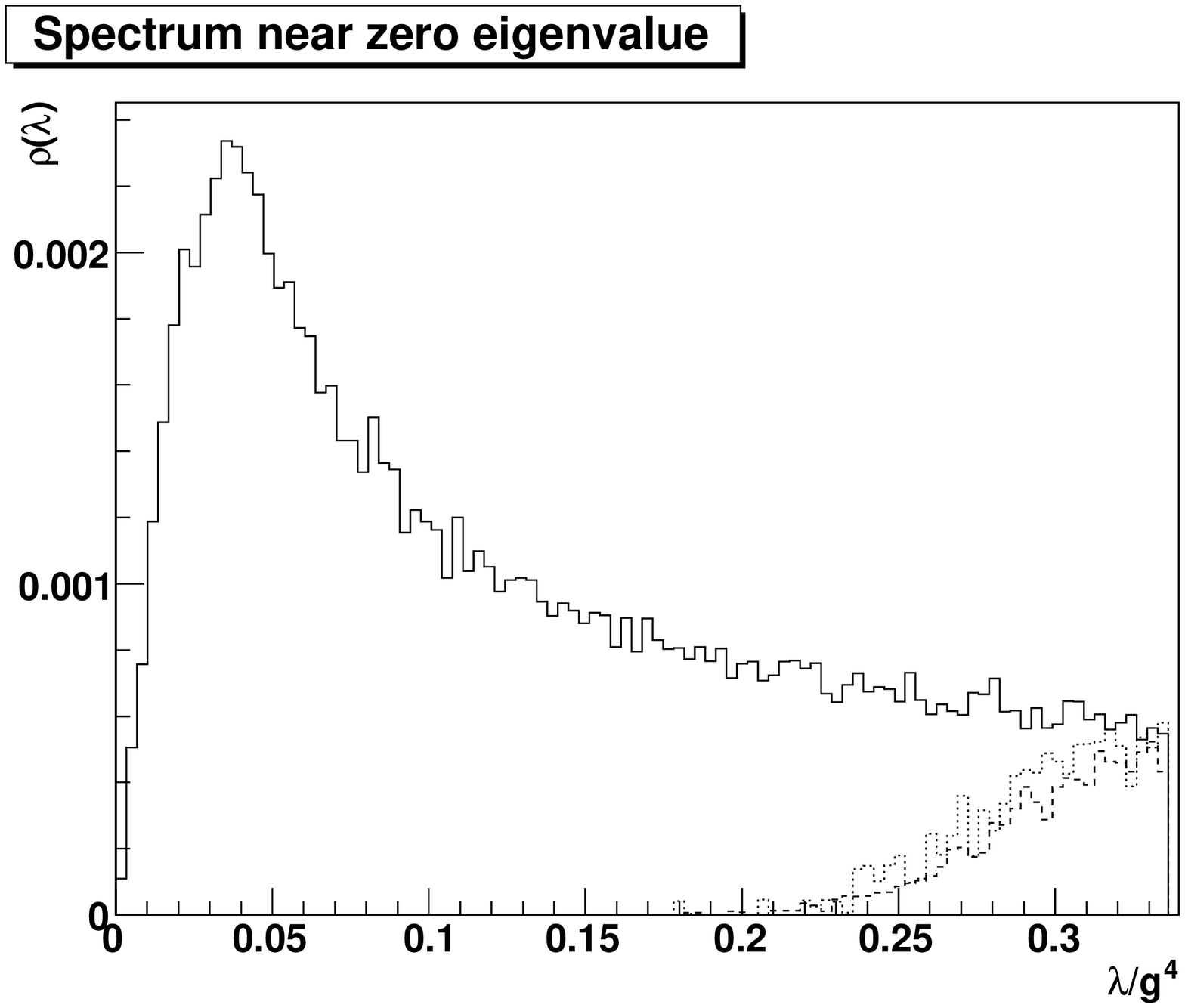}
\caption{\label{fles}The lower part of the eigenvalue spectrum of the Faddeev--Popov operator in two (left panel) and three (right panel) dimensions. Dotted lines are from gauge group G$_2$, dashed lines from SU(3), and solid lines from SU(2). This is an approximated spectrum, due to the numerical methods involved \cite{Cucchieri:2006tf}. 903747 (SU(2)), 4731736 (SU(3)), 843780 (G$_2$) and 576498 (SU(2)), 1057849 (SU(3)), and 275309 (G$_2$) eigenvalues have been evaluated in two and three dimensions, respectively for the whole spectrum. In the displayed part of the spectrum 20445, 12036, 6651, 57999, 9252, and 2952 eigenvalues are included, respectively. Note that the spiked structure is an artifact of binning and the amount of statistics. The lowest eigenvalue of the Laplacian in the same units are in two dimensions 0.085, 0.16, 0.12 and in three dimensions 0.17, 0.48, 0.27 for the gauge groups SU(2), SU(3), and G$_2$, respectively. The decrease in the spectrum of SU(2) towards large eigenvalues in three dimensions (which also occurs in all other cases at sufficiently large eigenvalue) is likely an artifact of the numerical method employed \cite{Cucchieri:2006tf}.}}

The results are shown in figure \ref{fghp}. The propagators do not differ significantly, and even the dressing functions are quite similar, for both dimensionalities. All diverge, but the divergence is less pronounced with increasing number of generators, for both dimensionalities. Nonetheless, in all cases a divergence is seen, in agreement with the GZKO scenario, although this divergence is still rather weak quantitatively in three dimensions, in particular for SU(3) and G$_2$. Larger volumes will be necessary for a more unambiguous result. In addition, the differences between the gauge groups SU(2) and SU(3) would be more pronounced than in four dimensions \cite{Cucchieri:2007zm}, if the results would be plotted in units of the string tension, while they are quite similar when plotted in units of the dimensionful coupling.

The divergence of the ghost propagator is also manifest in the low-lying eigenspectrum of the Faddeev--Popov operator, which is in all cases enhanced compared to that of the ordinary Laplacian. The latter has its lowest eigenvalue at $4\sin^2(\pi/L)$ in lattice units. This is shown in figure \ref{fles}. Although in all cases the eigenvalue spectrum extends below the one expected from a simple Laplacian, and thus the perturbative case, the distinction of the three gauge groups is nearly always marked, in particular in three dimensions. Only in case of G$_2$ in three dimensions, this enhancement is rather weak: The lowest eigenvalue found is just somewhat more than half the lowest eigenvalue of the Laplacian. The weak enhancement corresponds to the ghost propagator being least enhanced in this case.

While this enhancement has been repeatedly observed in SU(2) \cite{Cucchieri:2006tf,Maas:2007uv,Cucchieri:2007ta} and SU(3) \cite{Sternbeck:2005vs} Yang--Mills theory, it is interesting to observe it also in G$_2$ Yang--Mills theory. The more interesting since in SU(2) Yang--Mills theory a close connection between center vortices and the Faddeev--Popov operator in Landau as well as Coulomb gauge has been observed \cite{Greensite:2004bz,Langfeld:2002dd,Langfeld:2005kp,Maas:2005qt}. In particular, removing center vortices removed the low eigenspectrum enhancement of the operator and consequently the infrared enhancement of the ghost propagator.

Now\footnote{We are grateful to Jeff Greensite for inspiring comments on this topic.}, in G$_2$ there can only exist center vortices with a trivial flux, which, e.\ g., leads consistently in the center-vortex framework to a vanishing of the asymptotic string tension \cite{Greensite:2006sm}. In that respect, G$_2$ Yang--Mills theory is much more similar to QCD and to Yang--Mills--Higgs theory with a Higgs in the fundamental representation in the so-called confinement phase: In both cases, the center symmetry is broken, and string-breaking is accompanied by a low-eigenvalue enhancement of the Faddeev--Popov operator \cite{Fischer:rev,Ilgenfritz:2006he,Alkofer:2006gz,Fischer:2003rp,Bertle:2003pj,Greensite:2006ng}\footnote{Here it is anticipated that the low-eigenvalue enhanced eigenspectrum of the Coulomb-gauge Faddeev--Popov operator implies also an enhancement for the Landau-gauge Faddeev--Popov operator. This, of course, is still to be proven.}.

Therefore it seems likely that the same mechanism, which provides a coexistence of both phenomena in QCD and Yang--Mills--Higgs theory, is also at work in G$_2$ Yang--Mills theory. Even more so, as the charge structure of the gluons in G$_2$ Yang--Mills theory resembles closely the one of SU(3) Yang--Mills--Higgs theory and of QCD \cite{Holland:2003jy}. This would be in good agreement with the, very general, arguments in favor of an infrared enhanced Faddeev--Popov operator \cite{Maas:2005ym,Watson:2001yv}, aside from the GZKO mechanism.

Furthermore, the thermodynamic phase transition in SU($N$) Yang--Mills theory does break the center symmetry. However, the infrared properties of the propagators in Landau gauge, and, in particular, the low-lying spectrum of the Faddeev-Popov operator is not affected \cite{Cucchieri:2007ta}. Combining this with the findings on QCD and SU($N$)-Yang--Mills theory, as well as the results here on G$_2$ Yang--Mills theory, permits to conclude, or at least conjecture, more far-reaching consequences:  The presence or absence of a non-trivial center symmetry is related neither to the infrared behavior of the Landau gauge propagators nor to the Faddeev-Popov operator: Center symmetry, as relevant as it is to the linear rising potential, has seemingly no relevance for the GZKO mechanism.

However, vortices, independently of whether they carry a trivial or non-trivial center flux, are relevant to the GZKO mechanism, as various investigations have demonstrated explicitly \cite{Greensite:2004bz,Langfeld:2002dd,Langfeld:2005kp,Maas:2005qt}. As vortex field configurations, on the other hand, are likely the relevant carrier of the center symmetry \cite{Greensite:2003bk}, both aspects are not entirely unrelated, but there are quite subtle connections in SU($N$) Yang--Mills theory.

A necessary ingredient to permit such an argument is, of course, to observe a low-eigenvalue enhancement of the Faddeev--Popov operator in G$_2$ Yang--Mills theory at distance scales far larger than the string-breaking scale, which remains to be demonstrated.

Hence it is not surprising that the gluonic, and therefore the dynamical sector of Yang--Mills theory, seems to exhibit the same qualitative behavior independent of the gauge group, in accordance with previous conjectures \cite{Maas:2005ym}. In fact, the complete construction of the infrared behavior of Green's functions in SU($N$) gauge theory \cite{Huber:2007kc,Alkofer:2004it,Fischer:2006vf} can be carried over to arbitrary (semi-)simple gauge groups, as the color structure separates trivially, as long as only the tree-level color structure is present. No evidence to the contrary has been found so far to this, and also the results herein confirm it. However, the situation is more complicated when introducing quarks, as to let the string-tension vanish in G$_2$ QCD this scenario \cite{Alkofer:2006gz} requires subtle cancellations.

\section{Summary}\label{ssum}

In the present paper, a first determination of the ghost and gluon propagator, as well as of the low part of the eigenspectrum of the Faddeev--Popov operator, in G$_2$ Yang--Mills theory were presented. Although only in two and three dimensions and on rather small volumes, the comparison with SU(2) and SU(3) Yang--Mills theory has shown that no qualitative difference between these three gauge groups is present. 

In particular, at least in two dimensions, the explicit turnover of the gluon propagator has been demonstrated, as well as the infrared enhancement of the ghost propagator. Therefore the Gribov--Zwanziger and Kugo--Ojima mechanism for the removal of gluons from the physical spectrum seems to be at least as relevant to G$_2$ as to SU($N$) Yang--Mills theory. This is in agreement with previous expectations that the infrared properties of Yang--Mills theory should be independent of the gauge group for an arbitrary (semi-)simple Lie group \cite{Maas:2005ym}. Independent of this mechanism, the results also indicate that the gluon may be removed from the physical spectrum likely in the same way as in SU($N$) Yang--Mills theory.

However, it seems that quantitatively the characteristic infrared exponents or the scales in G$_2$ Yang--Mills theory seem to be different from those of SU(2) Yang--Mills theory. A more elaborate comparison along the lines of \cite{Maas:2007uv} will be necessary to decide whether the characteristic infrared exponents are in fact equal.

This is therefore another result, among those on QCD \cite{Fischer:rev,Ilgenfritz:2006he,Alkofer:2006gz,Fischer:2003rp}, Yang--Mills Higgs theory \cite{Bertle:2003pj,Greensite:2006ng}, and the phase transition in SU($N$) Yang--Mills theory \cite{Cucchieri:2007ta}, that the absence or presence of a non-trivial center symmetry is not relevant for the infrared behavior of (Landau-gauge) propagators or the spectral properties of the Faddeev--Popov operator. Therefore, this symmetry, as important as it is to the formation of a linear rising potential, is potentially unrelated to the GZKO mechanism. However, vortices, as field configurations, independently of their center charge, are still relevant to the GZKO mechanism \cite{Greensite:2004bz,Langfeld:2002dd,Langfeld:2005kp,Maas:2005qt}.

If this conjecture turns out to be correct, more light is shed on the subtle relations between the various non-perturbative aspects of Yang--Mills-type gauge theories.

\acknowledgments
Helpful discussions with Christian S.\ Fischer are acknowledged. The authors are also grateful to Jeff Greensite and Christian S.\ Fischer for critical reading of the manuscript and helpful comments. This work was supported by the DFG under grant MA 3935/1-2, and in part by 
the Slovak Science and Technology Assistance Agency under Contract 
No.\ APVT--51--005704, and by the Slovak Grant Agency for Science, Project VEGA No.\ 2/6068/2006. Additional computing time on the cluster of the theoretical particle physics group of the Karl-Franzens-Universit\"at Graz is highly acknowledged. The ROOT framework \cite{Brun:1997pa} has been used in this project.

\appendix

\section{Technical details}\label{atech}

\subsection{Overrelaxation in G$_2$}\label{aover}

Overrelaxation is performed by overrelaxing individual SU(2) subgroups. First, the staple matrix $k$ \cite{Montvay:1994cy} for a given link $U_\mu$ is determined. It is then changed by a similarity transformation to $k'=zkz^{-1}$, where $z$ is a random element of the coset group G$_2$/SU(3). Each matrix $A$ of G$_2$ can be written as \cite{Macfarlane:2002hr}
\be
A=Z\;\diag(V,1,V^*)=\begin{pmatrix}A & \vec a & B \cr \vec b^T & \alpha & \vec c^T \cr C & \vec d & D \cr\end{pmatrix}\label{notag2},
\ee
\no where $Z$ is an element of the coset group G$_2$/SU(3), and $V$ is an SU(3) element. The second equality is a general decomposition to fix notation, with $3\times3$ complex matrices $A$, $B$, $C$, and $D$, 3-dimensional complex vectors $\vec a$, $\vec b$, $\vec c$, and $\vec d$, and a complex scalar $\alpha$. The SU(2) subgroups are then isolated by determining the sum\footnote{Averaging is not necessary, as this quantity will be normalized later.} $A$+$D$ of $k'$. From this $3\times3$ matrix one SU(3) element is constructed in one of the SU(2) subgroups of SU(3) in the same manner as for SU(3) \cite{Cabibbo:1982zn}, yielding an SU(3) element $W$. The over-relaxed link $U_\mu'$ is then given by
\be
U_\mu'=z^{-1}\;\diag(W,1,W^*)\;z\;U_\mu.\nn
\ee
\no This is repeated for a given number (in the present case 3) subgroups of SU(3), each time with new mixing matrices $z$.

\subsection{Projection of matrices into G$_2$}\label{aproj}

In the course of gauge-fixing using stochastic overrelaxation it is necessary to perform a so-called Los-Alamos step \cite{Cucchieri:1995pn}. This leads to a general matrix $m_0$, which has to be projected into the appropriate group under the condition that the projected matrix $r$ maximizes the expression $\tr(r m_0)$ \cite{Suman:1993mg}. The iterative method to perform this in SU(3) \cite{Suman:1993mg}\footnote{Note that there are some misprints in the algorithm presented in \cite{Suman:1993mg}.} can be generalized to G$_2$ in the following way:

\begin{enumerate}
\item
Take some initial value for the matrix $r$, e.g. the unit matrix, and set a matrix $m$ to the value of the original matrix $m_0$. 
\item
Iterate the following algorithm:
\begin{enumerate}
\item
Select a random mixing matrix $m_x$ from G$_2$, and obtain the mixed matrix $m_x m m_x^{-1}$. 
\item
Using the notation introduced in \pref{notag2}, take any\footnote{Due to the mixing, it is irrelevant which. In the code used here, just $D$ was taken.} linear combination of $A$ and $D$. 
\item
Construct from it, according to the algorithm of \cite{Suman:1993mg}, an SU(3) matrix $E$. 
\item
Revert the similarity transformation to obtain $F=m_x^{-1} \diag(E,1,E^*) m_x$ and replace the matrix $r$ by $Fr$. 
\item
Repeat this sufficiently often (about ten to twenty times is usually more than sufficient) to cover the complete group.
\end{enumerate}
\item
Check whether $\tr(r m_0)$ has only changed within the desired precision. If yes, the algorithm has finished \cite{Suman:1993mg}, if not set $m=r m_0$, and repeat until it is the case.
\end{enumerate}

The amount of iterations needed depends on how well the gauge-fixing has already progressed. While more than a hundred iterations in the beginning do occur, in the end regularly below five are needed.

The gauge-fixing performance is not negatively affected, if in the early stages the required precision is less, or a limit on the number of iterations is set. However, quite high precision is needed to reach good accuracy in gauge-fixing. In fact, too little precision for the finishing condition can lead to a non-convergence of the stochastic overrelaxation. For the required precision in table \ref{tconf}, a precision of $10^{-16}$ and a maximum iteration limit below 400 was sufficient to guarantee convergence of the gauge-fixing in all cases.

\clearpage

\end{document}